\documentclass[fleqn,usenatbib,letters]{mnras}

\usepackage{newtxtext,newtxmath}

\usepackage[T1]{fontenc}

\usepackage{graphicx}	
\usepackage{amsmath}	
\usepackage{multirow}
\usepackage[table]{xcolor}
\usepackage{siunitx}

\title[Forecasts on axions, neutrinos and light elements]{Cosmological forecasts on thermal axions, relic neutrinos and light elements}

\author[W. Giarè, F. Renzi, A. Melchiorri, O. Mena and E. Di Valentino]{
William Giarè,$^{1}$\thanks{E-mail: william.giare@uniroma1.it }
Fabrizio Renzi,$^{2}$\thanks{E-mail: renzi@lorentz.leidenuniv.nl }
Alessandro Melchiorri,$^{1}$\thanks{E-mail: alessandro.melchiorri@roma1.infn.it}
Olga Mena,$^{3}$\thanks{E-mail: omena@ific.uv.es}
and Eleonora Di Valentino,$^{4}$\thanks{E-mail: eleonora.di-valentino@durham.ac.uk}
\\
$^{1}$Physics Department and INFN, Universit\`a di Roma ``La Sapienza'', Ple Aldo Moro 2, 00185, Rome, Italy\\
$^{2}$ Lorentz Institute for Theoretical Physics, Leiden University, PO Box 9506, Leiden 2300 RA, The Netherlands\\
$^{3}$ IFIC, Universidad de Valencia-CSIC, 46071, Valencia, Spain
\\
$^{4}$Institute for Particle Physics Phenomenology, Department of Physics, Durham University, Durham DH1 3LE, UK}

\pubyear{2021}

\begin{document}
\label{firstpage}
\pagerange{\pageref{firstpage}--\pageref{lastpage}}
\maketitle

\begin{abstract}
One of the targets of future Cosmic Microwave Background and Baryon Acoustic Oscillation measurements is to improve the current accuracy in the neutrino sector and reach a much better sensitivity on extra dark radiation in the Early Universe. In this paper we study how these improvements can be translated into constraining power for well motivated extensions of the Standard Model of elementary particles that involve axions thermalized before the quantum chromodynamics (QCD) phase transition by scatterings with gluons. Assuming a fiducial $\Lambda$CDM cosmological model, we simulate future data for Stage-IV CMB-like and Dark Energy Spectroscopic Instrument (DESI)-like surveys and analyze a mixed scenario of axion and neutrino hot dark matter. We further account also for the effects of these QCD axions on the light element abundances predicted by Big Bang Nucleosynthesis. The most constraining forecasted limits on the hot relic masses are  $m_{\rm a} \lesssim 0.92$ eV and $\sum m_\nu\lesssim 0.12$~eV at 95 per cent Confidence Level, showing that future cosmic observations can substantially improve the current bounds, supporting multi-messenger analyses of axion, neutrino and primordial light element properties.
\end{abstract}

\begin{keywords}
Cosmology: observations -- cosmic background radiation -- cosmological parameters -- dark matter -- early Universe
\end{keywords}



\section{Introduction} \label{sec.Introduction}
The most compelling solution to the strong CP problem in Quantum Chromodynamics (QCD) would require the Lagrangian of the Standard Model of elementary particles to be invariant under an additional global $U(1)_{\rm PQ}$ (Peccei-Quinn) symmetry, spontaneously broken at some energy scale $f_{\rm a}$~\citep{Weinberg:1975ui,Nanopoulos:1973wz,Weinberg:1973un,Belavin:1975fg,Callan:1976je,Jackiw:1976pf,Shifman:1979if,Kim:1979if,Dine:1981rt,Peccei:1988ci,Peccei:2006as,Peccei:1977hh,Peccei:1977ur,Wilczek:1977pj,Berezhiani:1989fp,Berezhiani:1992rk}. The result is an associated Pseudo Nambu Goldstone Boson (PNGB), the \emph{axion}~\citep{Weinberg:1977ma,Kim:1986ax,Shifman:1979if,Kim:1979if,Dine:1981rt,Cheng:1987gp,Peccei:1988ci,Peccei:2006as,Marsh:2015xka, DiLuzio:2020wdo,Sikivie:2006ni}. 

Axions could be copiously produced in the early universe both via thermal and non-thermal processes~\citep{Abbott:1982af,Dine:1982ah,Preskill:1982cy,Linde:1985yf,Seckel:1985tj,Lyth:1989pb,Linde:1990yj,Vilenkin:2000jqa,Kibble:1976sj,Kibble:1982dd,Vilenkin:1981kz,Davis:1986xc,Vilenkin:1982ks, Sikivie:1982qv,Vilenkin:1982ks,Huang:1985tt, Melchiorri:2007cd,Hannestad:2007dd,Hannestad:2008js,Hannestad:2010yi,Archidiacono:2013cha,Giusarma:2014zza,DiValentino:2015zta,DiValentino:2015wba,Archidiacono:2015mda,Hannestad:2005df}. Those produced non-thermally behave as Cold Dark Matter, while thermal axions (produced by interactions with other particles of the Standard Model) contribute to the Hot Dark Matter component of the Universe. In what follows we shall concentrate on the thermal axion scenario.

In the thermal production mechanism, axions can be described by two parameters: the axion coupling constant $f_a$ and the axion mass $m_a$, related as
\begin{equation}
m_{\rm a}=\frac{f_{\pi} m_{\pi}}{f_{a}} \frac{\sqrt{R}}{1+R}\simeq 0.6\, \mathrm{eV} \times \frac{10^{7}
	\,\mathrm{GeV}}{f_{a}}~,
\end{equation}
where $R\doteq m_u/m_d\simeq0.553 \pm 0.043 $ is the up-to-down quark mass ratio and $f_{\pi}\simeq 93$ MeV is the pion decay constant \citep{Zyla:2020zbs}. 

As along as thermal axions remain relativistic particles, they behave as extra dark radiation, contributing to the effective number of relativistic degrees of freedom $N_{\rm eff}$, modifying the damping tail of the CMB temperature angular power spectrum and changing two important scales at recombination: the sound horizon and the Silk damping scale. Furthermore, also the primordial abundances of light elements predicted by the Big Bang Nucleosynthesis (BBN) are sensitive to extra light species since larger values of $N_{\rm eff}$ would increase the expansion rate of the universe, leading to a higher freeze-out temperature for weak interactions and implying a higher fraction of primordial helium. 

Conversely, when thermal axions become non-relativistic particles, they leave cosmological signatures that are very similar to those originated by massive neutrinos. These axions therefore  contribute to the (hot) dark matter component, suppressing structure formation at scales smaller than their free-streaming scale and leaving an imprint on the CMB temperature anisotropies, via the early integrated Sachs-Wolfe effect. So, a large degeneracy between the axion and the total neutrino masses is typically expected. 

Updated bounds on thermal axion masses should be then obtained within a realistic mixed hot dark matter scenario, including also massive neutrinos\footnote{As robustly indicated by oscillation experiments~\citep{deSalas:2020pgw,deSalas:2018bym}, neutrinos should be regarded as massive particles and cosmology provides a powerful (albeit indirect) mean to constrain their mass~\citep{deSalas:2018bym,Hagstotz:2020ukm,Vagnozzi:2019utt,Vagnozzi:2018pwo,Vagnozzi:2018jhn,Vagnozzi:2017ovm,Giusarma:2016phn,Bond:1980ha,Doroshkevich:1980zs}. However, a quite strong correlation between $\sum m_{\nu}$ and $H_0$ is typically observed with possible implications for the cosmological tensions~\citep{Capozzi:2021fjo,DiValentino:2021imh}.}. For the thermal axions decoupling before the QCD phase transition, the most constraining bounds are $m_{\rm a}<7.46$ eV and $\sum m_{\nu}<0.114$ eV both at 95\%CL~\citep{Giare:2020vzo}. 
For axions decoupling after the QCD phase transition, these bounds can be improved to $m_a<0.91$~eV and $\sum m_\nu< 0.105$~eV, both at 95\% CL\footnote{Notice anyway that in the latter case the upper bound on the axion mass is mostly due to the limitation of the range of validity of chiral perturbation theory and weakly depend on data sensitivity}. These limits were obtained by means of the final release of Planck 2018 temperature and polarization data~\citep{Akrami:2018vks} in combination with the other most recent (CMB-independent) cosmological observations. While these bounds are able to probe a significant range of the parameter space allowed by direct axion searches, the current constraining power on the total variation of the effective number of relativistic species due to extra dark radiation ($\Delta N_{\rm eff} \lesssim 0.4$ at 95\% CL) represents an important limitation as it is not enough accurate to reveal the presence of thermal axions produced before the QCD transition. These axions would lead to $\Delta N_{\rm eff}\lesssim 0.1$, which lies well below the present sensitivity to $\Delta N_{\rm eff}$. In this regard, the next generation of CMB experiments is expected to significantly increase the constraints on $N_{\rm eff}$. In particular, a Stage-IV CMB experiment (CMB-S4) will increase the accuracy on extra dark radiation by almost an order of magnitude, $\Delta N_{\rm eff} \lesssim 0.06$ \citep{Abazajian:2016yjj}, opening to the possibility of robustly constrain the mass of QCD axions also for this thermal channel. In addition, future observations of large scale structure will lead to highly accurate Baryon Acoustic Oscillation (BAO) data, expected to provide an enormous improvement on the neutrino mass bound~\citep{DESI}.  Interestingly, the combination of CMB-S4 and Dark Energy Spectroscopic Instrument (DESI) should be able to constrain the sum of the neutrino masses at the $2\sigma$ level with a precision of $\sigma(\sum m_\nu) \sim 16\ $meV \citep{constraint_neutrino_stageIV,Xu:2020fyg} for $m_{\nu} \sim 0.06 $ eV.

In this work, focusing exclusively on QCD axions produced \textit{before} the QCD phase transition, we analyze realistic mixed hot dark matter scenarios that include also massive neutrinos. The aim is to study the improvement in the constraining power on hot relics expected by the next-gen CMB and BAO observations. We also investigate and discuss the implications for BBN light element primordial abundances up to Beryllium-7. 

The paper is organized as follows. In \autoref{sec.Theory} we review the thermalization processes involved in the axion production; in \autoref{Sec.Method} we describe the numerical methods adopted to simulate the data used for the forecasts; in \autoref{Sec.Results} we present and discuss our results; finally, in \autoref{Sec.Conclusions} we present our conclusions.

\section{Axion thermalization}\label{sec.Theory}
The axion contribution to the effective number of relativistic degrees of freedom 
\begin{equation}
 \Delta N_{\rm eff}  \simeq \frac{4}{7} \left[ \frac{43}{4\,g_{\star\,S}(T_{d})}\right]^{4/3},
\end{equation}
depends on the temperature $T_d$ at which axions decouple from the thermal bath via the number of entropic degrees of freedom $g_{\star\,S}(T)$. Notice that the decoupling temperature can be estimated solving the usual freeze-out condition
\begin{equation}
H(T_d)=\Gamma(T_{d})~,   
\end{equation}
where $\Gamma(T)$ is the axion interaction rate and $H(T)$ is the expansion rate, which, in a radiation dominated Universe, is given by
\begin{equation}
H(T_d)=\sqrt{\frac{4 \pi^{3}}{45} g_{\star}(T_d)} \left(\frac{T_d^{2}}{M_{p l}}\right),
\end{equation}
with $M_{pl}\simeq 1.22 \times 10^{19} \rm{GeV}$ the Planck Mass and $g_{\star}(T_d)=g_{\star}^{\rm SM}(T_d)+1$ the Standard Model relativistic degrees of freedom with the additional contribution from the axion. On the other hand, considering only the two-body processes with cross sections $\sigma_i = \sigma(p_i a \leftrightarrow p_jp_k)$ and with all the particles in thermal equilibrium, the axion interaction rate reads~\citep{Melchiorri:2007cd}
\begin{equation}
\Gamma \propto \sum_{i} n_{i}\left\langle v \sigma_{i}\right\rangle
\end{equation}
where $n_i$ is the number density of $p_i$, $v\simeq 1$ the relativistic velocity and the brackets refer to the thermal average. 

After decoupling ($T<T_d$), axions maintain a thermal distribution which basically remains unaffected by other phenomena occurring in the plasma and therefore the current axion number density is simply given by
\begin{equation}
n_{a}=\frac{g_{\star S}\left(T_{0}\right)}{g_{\star S}\left(T_{d}\right)} \times \frac{n_{\gamma}}{2}~,
\label{na}
\end{equation}
with $n_{\gamma}\simeq 411 \mathrm{~cm}^{-3}$ the present photon density and $g_{\star S}(T_0)\simeq 3.91$ the current number of entropic degrees of freedom (we recall that  before the neutrino freeze-out $g_{\star S}=g_{\star}$).

In the Early Universe, there are several processes that can keep the axion in thermal equilibrium with the Standard Model thermal bath and, depending on whether axions decouple before or after the QCD phase transition ($T_{\rm QCD}\simeq 160\,\rm{MeV}$), different thermal channels should be considered~\citep{Ferreira:2018vjj,DEramo:2021usm,DEramo:2021lgb,DEramo:2021psx,Arias-Aragon:2020shv,DEramo:2018vss,Ferreira:2020bpb}.  If axions decouple before the QCD phase transition ($T_d>T_{\rm QCD}$) the most relevant thermalization channel is provided by the axion-gluon scattering~\citep{Melchiorri:2007cd,Hannestad:2007dd,Hannestad:2008js,Hannestad:2010yi,Archidiacono:2013cha,Giusarma:2014zza,DiValentino:2015zta,DiValentino:2015wba,Archidiacono:2015mda,Hannestad:2005df}. Indeed in any QCD axion model, axions necessarily couple to gluons via a model-independent coupling $\alpha_s/(8\pi\,f_a) a\,G\tilde G$ required to solve the strong CP problem. In this case the relevant processes for axion thermalization are:
\begin{itemize}
\item $a + q \leftrightarrow g  + q$ and  $a  + \bar{q} \leftrightarrow g + \bar{q}$;
\item $a + g \leftrightarrow q + \bar{q}$;
\item $a + g \leftrightarrow g + g$. 
\end{itemize}
and the decoupling temperature for this axion population is estimated to be~\citep{DiLuzio:2020wdo} 
\begin{equation}
T_{d} \simeq 12.5 \frac{\sqrt{g_{*}(T)}}{\alpha_{s}^{3}} \frac{f_{a}^{2}}{M_{pl}} \mathrm{GeV}.
\label{Td_fa}
\end{equation}

If instead axions decouple after the QCD phase transition ($T_d<T_{\rm QCD}$), processes involving pions and nucleons must be considered. In practice, however, nucleons are so rare in the early universe with respect to pions that the only relevant process is the axion-pion interaction $\pi+\pi \leftrightarrow \pi + a $. In this case, a leading order computation in chiral perturbation theory predicts the interaction rate~\citep{DiLuzio:2020wdo, DiLuzio:2021vjd}
\begin{equation}
\Gamma_{a \pi}^{\rm LO} \simeq 0.215\, C_{a \pi}^{2}\, \frac{T^{5}}{f_{a}^{2} f_{\pi}^{2}}\, h_{\rm{LO}}\left(\frac{m_{\pi}}{T}\right),
\label{Gamma_api}
\end{equation}
where $h(x)$ is a rapidly decreasing function of its arguments (and normalized to $h(0)=1$), and $C_{a \pi}$ is the axion-pion coupling, which is given by 
\begin{equation}
C_{a \pi}=\frac{1}{3}\left(\frac{m_{d}-m_{u}}{m_{u}+m_{d}}+c_{d}^{0}-c_{u}^{0}\right)~.
\end{equation}
The axion-pion coupling is a model-dependent quantity, sensitive to the nature of axion-fermion interactions via the axion-quark couplings $c_{d}^{0}$ and $c_{u}^{0}$. Therefore the thermal production of axions via pion scattering is strongly model-dependent since the relation between the axion mass and the (decoupling) temperature changes accordingly to the axion-pion interaction strength and consequently, the thermal production could range between relatively large thermal abundances to negligible ones, depending on the precise value of $C_{a \pi}$. Furthermore, the perturbative approach can only be extended up to temperatures $T_{\rm d}\lesssim 62\,\rm{MeV}$ since above this limit the next-to-leading order corrections become of the same order than the leading order part ($\Gamma_{a\pi}^{\rm NLO}\simeq \Gamma^{\rm L0}_{a\pi}$)~\citep{DiLuzio:2021vjd}. Consequently, no reliable bounds on axion masses can be derived until robust lattice QCD techniques would provide the precise answer for any given model in these temperature ranges. In this work we ignore this model-dependent thermal channel and we exclusively focus on the axion-gluon scatterings.

\section{Method}\label{Sec.Method}
In this section we describe the method followed for our forecasted analyses, focusing on future CMB and BAO observations. In particular, we simulate future data for a CMB-S4-like \citep{Abazajian:2016yjj} observatory and for a DESI-like \citep{DESI,DESI:2013agm} BAO survey. These probes are expected to provide scientific results in the next few years and have been carefully designed to improve the constraints on the neutrino sector and other forms of dark radiation in a significant way~\citep{Abazajian:2016yjj,DESI,constraint_neutrino_stageIV,Abazajian:2019oqj}. Finally, we also address the effect of additional thermal species on the observational prediction of BBN on light element abundances up to Beryllium-7.

All our forecasted datasets  make use of the  \textsc{COBAYA} software~\citep{Torrado:2020xyz}. The code allows to build synthetic realization of cosmological data for both CMB and BAO observations and test them again a given cosmological model. The parameter posteriors have been sampled using the MCMC algorithm developed for CosmoMC \citep{Lewis:2002ah,Lewis:2013hha}.
The predictions of the theoretical observational probes are calculated using the latest version of the cosmological Boltzmann integrator code \textsc{CAMB} \citep{Lewis:1999bs,Howlett:2012mh}. To include the effect of the axion-gluon coupling as an additional form of dark radiation we have modified the \textsc{CAMB} package accordingly to the description detailed in the previous section. The strength of the coupling and its effect on the neutrino sector are function only of the axion mass that we include as an additional cosmological parameter in our analyses. 

To complete this picture one needs to choose a fiducial cosmological model to build the forecasted data. We perform our forecasts using values of the parameters that are in agreement wit the latest Planck 2018 constraints for a $\Lambda$CDM scenario~\citep{Aghanim:2018eyx}. In particular we choose the following values for the standard six cosmological parameters :  $n_s = 0.965$, 
$\omega_b = 0.0224$, $\omega_c = 0.12$, $H_0 = 67.4$, $\tau=0.05$, $A_s = 2.1 \times 10^{-9}$, $N_{\rm eff} = 3.046$, $\sum m_\nu = 0.06$ eV and $m_{\rm a}=0$~eV. The above values are those commonly used in the forecasts available in the literature, and, therefore, for the sake of comparison, are the most convenient and useful ones, despite the fact that none of these previous works have considered $m_{\rm a}$ as a parameter to be constrained.

\subsection{CMB-S4 forecasts}
We build our forecast for future CMB observations using a well-established and robust method that is now a common practice in cosmological analyses. Using the fiducial model introduced above, we compute the angular power spectra of temperature $ C_\ell^{TT} $, E and B polarization $ C_\ell^{EE,BB} $ and cross temperature-polarization $C_\ell^{TE} $ anisotropies. Then, we consider an experimental noise for the temperature angular spectra of the form \citep{Perotto:2006rj}:
\begin{equation}
	N_\ell = w^{-1}\exp(\ell(\ell+1)\theta^2/8 \ln 2)~,
\end{equation}  
where $ \theta $ is the FWHM angular resolution and $ w^{-1} $ is the experimental sensitivity in units of $ \mu\mathrm{K}\,\rm arcmin $. The polarization noise is derived assuming $ w_p^{-1} = 2w^{-1} $ (one detector measures two polarization states). 
The simulated spectra are compared with theoretical ones using the following likelihood $\mathcal{L} $ \citep{Perotto:2006rj,Cabass:2015jwe}:
\begin{equation}
	-2\ln\mathcal{L}_{\rm CMB} = \sum_{\ell} (2\ell + 1)f_{\rm sky}\left(\frac{D_\ell}{|C_\ell|} + \ln\frac{|C_\ell|}{|\hat{C_\ell}|} - 3 \right)~,
\end{equation} 
where $\hat{C} $ and $ C $ are the theoretical and simulated spectra (plus noise), respectively and are defined by :
\begin{align}
&|C_\ell
| = C_\ell^{TT}C_\ell^{EE}C_\ell^{BB} -
\left(C_\ell^{TE}\right)^2C_\ell^{BB}~;  \\
&|\hat{C}_\ell| = \hat{C}_\ell^{TT}\hat{C}_\ell^{EE}\hat{C}_\ell^{BB} -
\left(\hat{C}_\ell^{TE}\right)^2\hat{C}_\ell^{BB}~,
\end{align}
while $ D $ is :
\begin{align}
	D_\ell  &=
	\hat{C}_\ell^{TT}C_\ell^{EE}C_\ell^{BB} +
	C_\ell^{TT}\hat{C}_\ell^{EE}C_\ell^{BB} +
	C_\ell^{TT}C_\ell^{EE}\hat{C}_\ell^{BB} \nonumber\\
	&- C_\ell^{TE}\left(C_\ell^{TE}\hat{C}_\ell^{BB} +
	2C_\ell^{TE}C_\ell^{BB} \right)~. \nonumber\\
\end{align}

In this study we construct synthetic realizations of CMB data for only one experimental configuration, namely CMB-S4 (see e.g. \citep{Abazajian:2016yjj}), using $ \theta = \SI{3}{\arcminute} $ and $ w = 1\, \si{\mu\kelvin}\,\rm arcmin $. The range of multipoles  is $ 5 \leq \ell \leq 3000 $ and the sky coverage of the $ 40\% $ ($f_{\rm sky} = 0.4$). We do not include CMB lensing derived from trispectrum data.

\subsection{DESI (BAO) forecasts}
For the future BAO dataset we consider the DESI experiment~\citep{DESI:2013agm}. As a tracer for BAO observations we employ the volume average distance defined as:

\begin{equation}
D_V(z)\equiv \left(\frac{(1+z)^2D_A(z)^2cz}{H(z)}\right)^\frac{1}{3}~,
\end{equation}

\noindent where $D_A$ is the angular diameter distance and $H(z)$ the Hubble parameter.
Assuming the fiducial model described previously, we compute the theoretical values of the ratio $D_V/r_s$ for several redshifts in the range $z=[0.15-1.85]$, where $r_s$ is sound horizon at the photon-baryon decoupling epoch. The uncertainties on $D_V/r_s$ are calculated propagating those for $D_A/r_s$ and $H(z)$ reported in \citep{DESI}. 
The simulated BAO data are compared to the theoretical $D_V/r_s$ values through a multivariate Gaussian likelihood :
\begin{equation}
    -2\ln \mathcal{L}_{\rm BAO} = \sum (\mathbf{\mu} - \hat{\mathbf{\mu}})C^{-1}(\mathbf{\mu} - \hat{\mathbf{\mu}})^T ~,
\end{equation}
where $\mu$ and $\hat{\mu}$ are the vectors containing the simulated and theoretical values of $D_V/r_s$ at each redshift and $C$ their simulated covariance matrix. 

\begin{figure}
    \centering
    \includegraphics[width=\columnwidth,keepaspectratio]{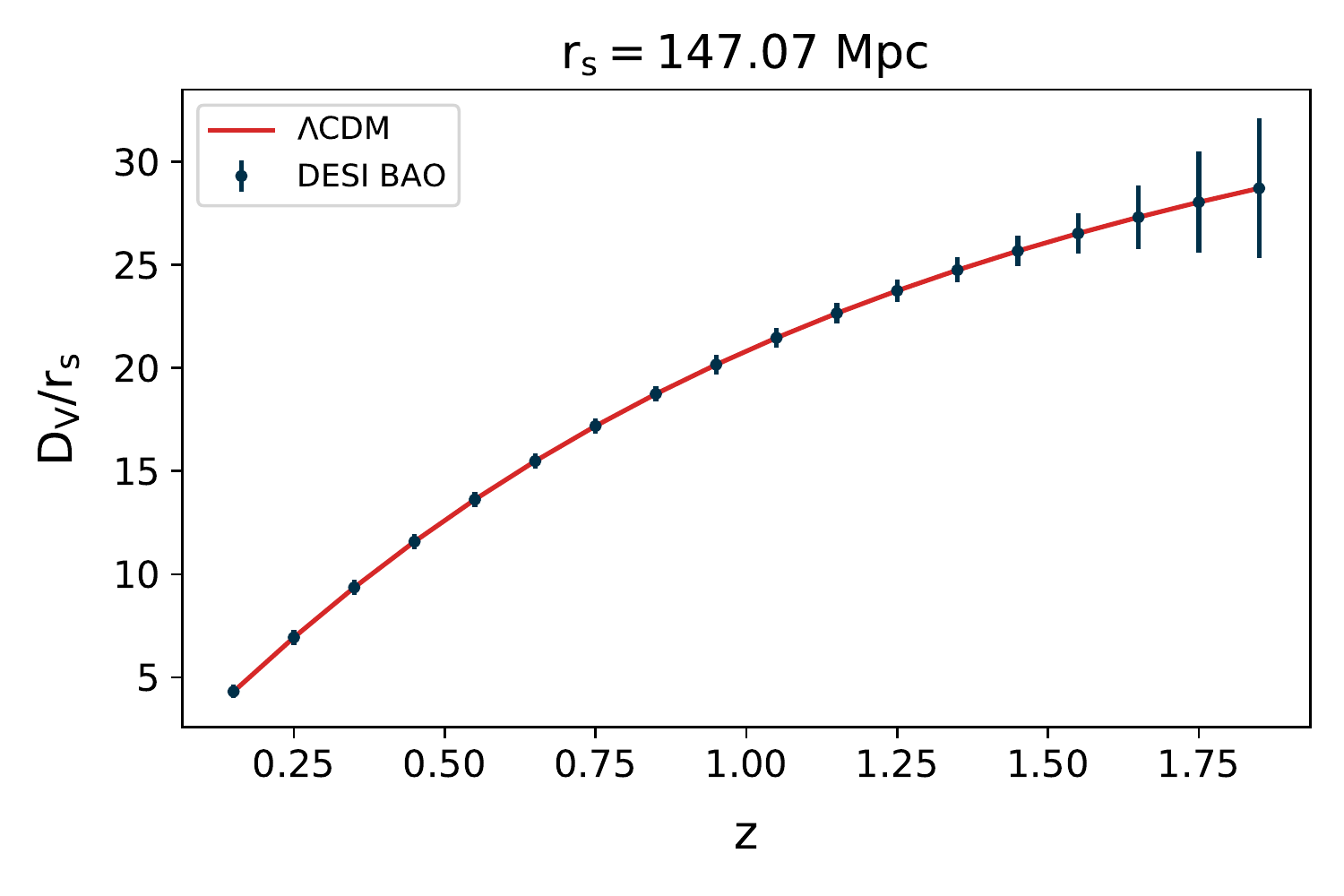}
\caption{The fiducial BAO datasets employed in our forecasts. Error bars refer to $3\sigma$ CL uncertainties.}
\label{fig:BAO_fid}
\end{figure}

It would also be possible to forecast BAO data considering $D_A/r_s$ and $H(z)$ as independent measurements, allowing for stronger constraints. However some small tension ($\sim 1$ sigma) has been identified between the current constraints from $D_A/r_s$ and $H(z)$ \citep{Addison:2017fdm}. Given the difficulty of properly accounting for this small tension between $D_A/r_s$ and $H(z)$ , we decided to follow the approach of \citep{Allison:2015qca} and employ the volume average distance for our BAO forecasts. The resulting dataset is the same obtained in Ref.~\citep{DiValentino:2018jbh} (see also their Table 2) while a plot representing the forecasted dataset is presented in \autoref{fig:BAO_fid}.

\subsection{BBN primordial light element abundances}

 Big Bang Nucleosynthesis (BBN) is a cornerstone of the Hot Big Bang cosmology which explains the formation of the first light nuclei (from $\rm H$ up to $^7\rm Li$) by a solid understanding of the nuclear interactions involved in the production of elements.

The set of differential equations that regulate those interactions in the primordial plasma can be solved numerically~\citep{PitrouEtal2018,Pisanti:2007hk,Consiglio:2017pot,Gariazzo:2021iiu} after neutrino decoupling ($T \gtrsim 1$ MeV) up the end of BBN ($T \sim 10$ keV), yielding the total abundance of primordial elements in terms of their ratio with respect to the hydrogen abundance. BBN provides a natural laboratory to probe new physical scenarios of the early Universe and its predictions can be compared to the primordial abundances of light elements inferred by astrophysical and cosmological observations. Given current uncertainties, BBN predictions and primordial light element measurements show a good agreement~\citep{Pitrou:2020etk,Mossa:2020gjc,Pisanti:2020efz} \footnote{Nevertheless, there are some discrepancies in the observed abundance of $^7\rm Li$ (which is a factor of $\sim 2$ smaller than those measured from low-metallicity stars~\citep{Fields:2011zzb,Cyburt:2008kw}) and in that  of the primordial deuterium (which exhibit a 1.8$\sigma$ discrepancy with the CMB+BAO value \citep{Pitrou:2020etk}).}. Notice also that since the BBN epoch ends before recombination, its outcome does not have any impact on the recombination epoch or else on the CMB power spectra. In other words, recombination and BBN are two independent and complementary probes that can be combined to check the consistency of particles interactions in the early universe.
What is relevant for CMB is the BBN prediction for the Helium abundance , that can be used to estimate the baryon density through a simple formula:
\begin{equation}
    \Omega_b h^2 = \frac{1 - 0.007125\ Y_p^{\rm BBN}}{273.279}\left(\frac{T_{\rm CMB}}{2.7255\ \mathrm{K}}\right)^3 \eta_{10}~, 
\end{equation}
where $\eta_{10} \doteq 10^{10}n_b/n_\gamma $ is the photon-baryon ratio today, $T_{\rm CMB} $ is the CMB temperature at the present time and $Y_p^{\rm BBN} \doteq 4 n_{\rm He}/n_{b}$ is the helium nucleon fraction defined as the ratio of the 4-Helium to the baryon density one.
Furthermore, BBN depends on the expansion rate $H(z)$, which sets the value of the temperature of the Universe during the radiation epoch via a function of the radiation density:
\begin{align}
&    H(z) \simeq \frac{8\pi G}{3}\rho_{\rm rad} \simeq \frac{7\pi G}{3} \,  N_{\rm eff} \left(\frac{4}{11}\right)^{4/3}\rho_\gamma~,
\end{align}
making BBN a very powerful tool to constrain the total number of relativistic species via $H(z)$. 

In this work, we made use of the code \textsc{PArthENoPE}~\citep{Gariazzo:2021iiu}. Using the values of $N_\mathrm{eff}$, $\tau_n$ (the neutron lifetime~\footnote{It is worth noting that, the interaction rates used in BBN codes assume a prior knowledge of $\tau_n$, which sets the efficiency of nuclear reactions. Therefore, BBN abundances are significantly affected even by a small change in the precise value of this parameter.}) and $\eta_{10}$ (or equivalently $\Omega_b h^2$), the code computes the value of $Y_P^{\rm BBN}$ and other light element abundances. To include the BBN code predictions in our MCMC analysis, we follow the same procedure used by the Planck collaboration~\citep{Aghanim:2018eyx}. Namely, we fix the neutron life time and create an interpolation grid varying $\Omega_b h^2 $ and $\Delta N_\mathrm{eff} = N_{\rm eff} - 3.045$ within a given range. We choose these ranges to be $\Delta N_\mathrm{eff} \in [-3 ; 3]$ and $\Omega_b h^2 \in [0.0073 ; 0.033]$. The neutron lifetime is fixed to $\tau_n = 879.4$ s, corresponding to the latest measurement reported by the Particle Data Group ($\tau_n = 879.4 \pm 0.6 $ s)~\citep{ParticleDataGroup:2020ssz} \footnote{This estimate of the neutron life time is derived averaging over a large number of measurements. However, beam-only and bottle-only experiments show a 4$\sigma $ discrepancy in measuring the neutron lifetime, leading respectively to $\tau_n = 888.0 \pm 2.0 $ s  and $\tau_n = 879.2 \pm 0.6$~s  (see also the discussion in Ref.~\citep{Salvati:2015wxa}). Interestingly, independent constraints can be derived by  CMB data only, but these limits are not accurate enough to disentangle between the two results ($\tau_n = 851 \pm 60$ s). Nevertheless, the neutron lifetime discrepancy is beyond the scope of this work and we therefore fix its value to that of Ref.~\citep{ParticleDataGroup:2020ssz}, even if this could produce a systematic error in $N_\mathrm{eff}$~\citep{Capparelli:2017tyx}. Other theoretical uncertenties that can affect the axion contribution to the relativiastic degrees of freedom in the early Universe are briefly discussed in \hyperref[AppendixA]{Appendix A}.}.

\begin{table*}
	\begin{center}
		\renewcommand{\arraystretch}{1.5}
		\begin{tabular}{c@{\hspace{1.5 cm}} c @{\hspace{1.5 cm}} c @{\hspace{1.5 cm}} c}
			\hline
			\textbf{Parameter}    &\textbf{Fiducial value} & \textbf{CMB-S4} & \textbf{CMB-S4 + DESI}\\
			\hline\hline
			$\Omega_{\rm b} h^2$      &$0.0224$    & $0.022420\pm 0.000034$  & $0.022422\pm 0.000034$\\
			$\Omega_{\rm c} h^2$      &$0.12$	 & $0.12066\pm 0.00062$  & $0.12019\pm 0.00032$\\
			$H_0$ [Km/s/Mpc]   &$67.4$  & $66.94^{+0.63}_{-0.57}$  & $67.39\pm 0.24$\\
			$\tau$                      &$0.05$  & $0.0508\pm 0.0026$  & $0.0508\pm 0.0025$\\
			$\log(10^{10}A_{\rm S})$   &$3.044$   & $3.0491\pm 0.0049$  & $3.0478\pm 0.0047$\\
			$n_{\rm s}$                  &$0.965$ & $0.9647\pm 0.0021$  & $0.9659\pm 0.0017$\\
			$\sum m_{\nu}$  [eV]         &$0.06$     & $< 0.183$  & $< 0.122$\\ 
			$m_{\rm a}$ [eV]                    &$0.0$     & $<1.60$  & $< 0.924$\\
			\hline 
			$Y_P^\mathrm{BBN}$           &$0.247$ & $0.247268^{+0.000052}_{-0.000085}$  & $0.247244^{+0.000030}_{-0.000042}$\\
			$10^5 \,\mathrm{D}/\mathrm{H}$ &$2.514$ & $2.5211\pm 0.0065 $  & $2.5200\pm 0.0063$\\
			$10^7 \, \mathrm{T}/\mathrm{H}$ &$0.808$ & $0.8104\pm 0.0022$  & $0.8101\pm 0.0021$\\
			$10^5 \,\mathrm{He3}/\mathrm{H}$  &$1.032$ & $1.03374\pm 0.00095$  & $1.03358\pm 0.00094$\\
			$10^{10}\, \mathrm{Li7}/\mathrm{H}$  &$4.67$ & $4.670\pm 0.015$  & $4.671\pm 0.015$\\
			$10^{10}\, \mathrm{Be7}/\mathrm{H}$  &$4.40$ & $4.396\pm 0.031$  & $4.398\pm 0.015$\\
			\hline\hline
		\end{tabular}
		\caption{Results for the $\Lambda\rm{CDM} + m_a + \sum m_{\nu}$ cosmological model and on the primordial light element adundances. The constraints on the parameters are at $68\%$ CL, while upper bounds are quoted at $95\%$ CL. We make use of the  \textsc{PArthENoPE} package to compute the BBN predictions.}
		\label{tab.Results.HDM}
	\end{center}
\end{table*}

\begin{figure*}
    \centering
    \includegraphics[width=0.8\textwidth]{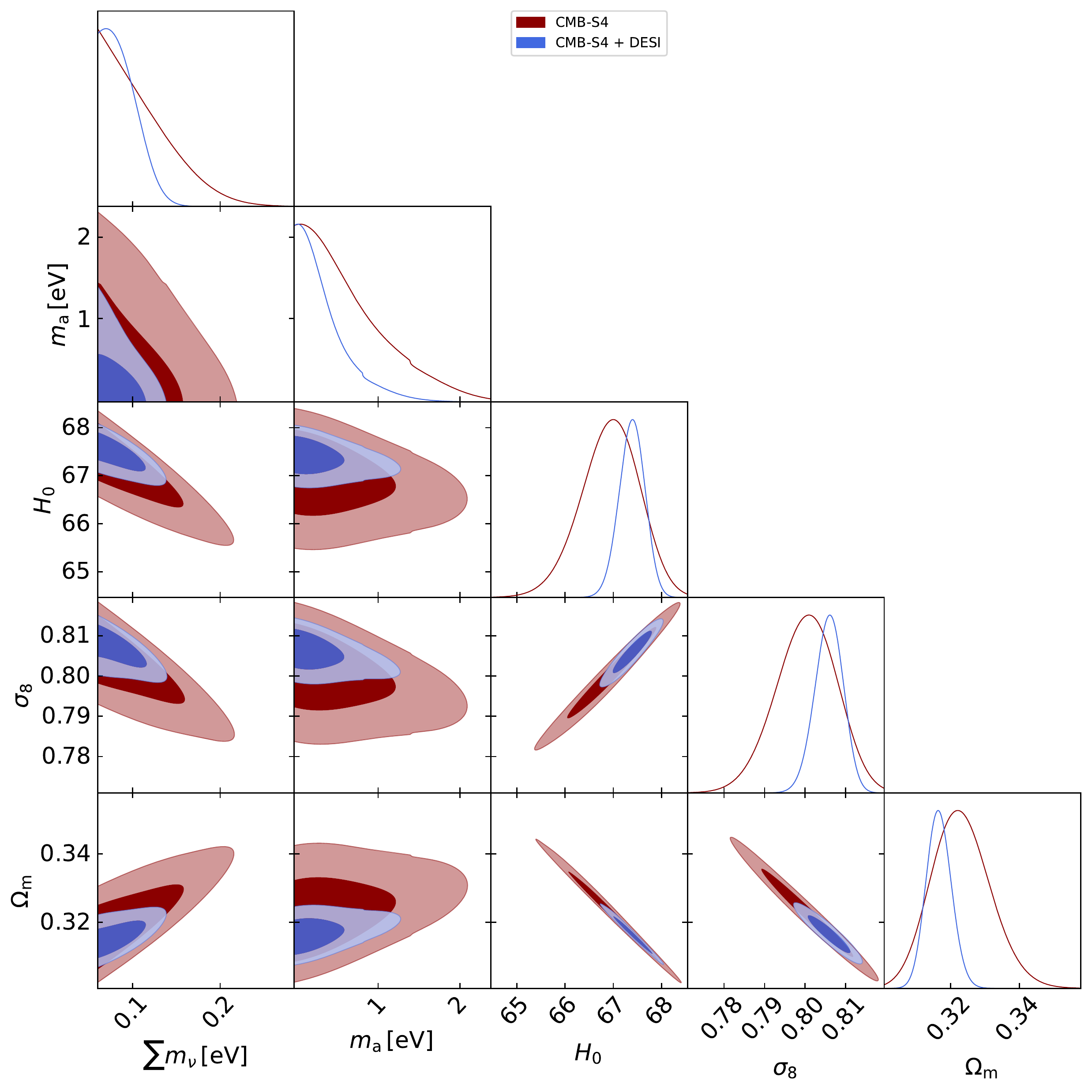}
\caption{Marginalized 2D and 1D posteriors for different cosmological parameters obtained from the forecasting data and methods.}
\label{fig:LCDM+ma+mnu}
\end{figure*}

\begin{table*}
	\begin{center}
		\renewcommand{\arraystretch}{1.5}
		\begin{tabular}{c@{\hspace{1.5 cm}} c @{\hspace{1.5 cm}} c @{\hspace{1.5 cm}} c}
			\hline
			\textbf{Parameter}    &\textbf{Fiducial value} & \textbf{CMB-S4} & \textbf{CMB-S4 + DESI}\\
			\hline\hline
			$\Omega_{\rm b} h^2$      &$0.0224$   & $0.022399\pm 0.000050$  & $0.022406\pm 0.000050$\\
			$\Omega_{\rm c} h^2$     &$0.12$	 & $0.12070\pm 0.00062$  & $0.12020\pm 0.00031$\\
			$H_0$ [Km/s/Mpc]   &$67.4$  & $66.90^{+0.65}_{-0.57}$  & $67.37\pm 0.24$\\
			$\tau$                     &$0.05$  & $0.0506\pm 0.0026$  & $0.0507\pm 0.0025$\\
			$\log(10^{10}A_{\rm S})$    &$3.044$ & $3.0486\pm 0.0049$  & $3.0474\pm 0.0047$\\
			$n_{\rm s}$                 &$0.965$ & $0.9635\pm 0.0034$  & $0.9649\pm 0.0030$\\
			$\sum m_{\nu}$ [eV]           &$0.06$    & $< 0.183$  & $< 0.120$\\ 
			$m_{\rm a}$ [eV]                    &$0.0$    & $ < 1.63$  & $< 0.991$\\
			$Y_p^{\rm BBN}$                  &$0.247$      &$0.2458^{+0.0057}_{-0.0058}$ & $0.2460^{+0.0057}_{-0.0058} $\\
			\hline\hline
		\end{tabular}
		\caption{Results for $\Lambda\rm{CDM} + m_{\rm a}+ \sum m_{\nu} + Y_p^{\rm BBN}$ case (i.e. leaving the Helium nucleon fraction as a free parameter of the model, without assuming the BBN theoretical predictions). The constraints on parameters are at $68\%$ CL, while the quoted upper bounds are at $95\%$ CL.}
		\label{tab.Results.LCDM+ma+mnu+Yp}
	\end{center}
\end{table*}

\begin{figure*}
    \centering
    \includegraphics[width=0.75\textwidth]{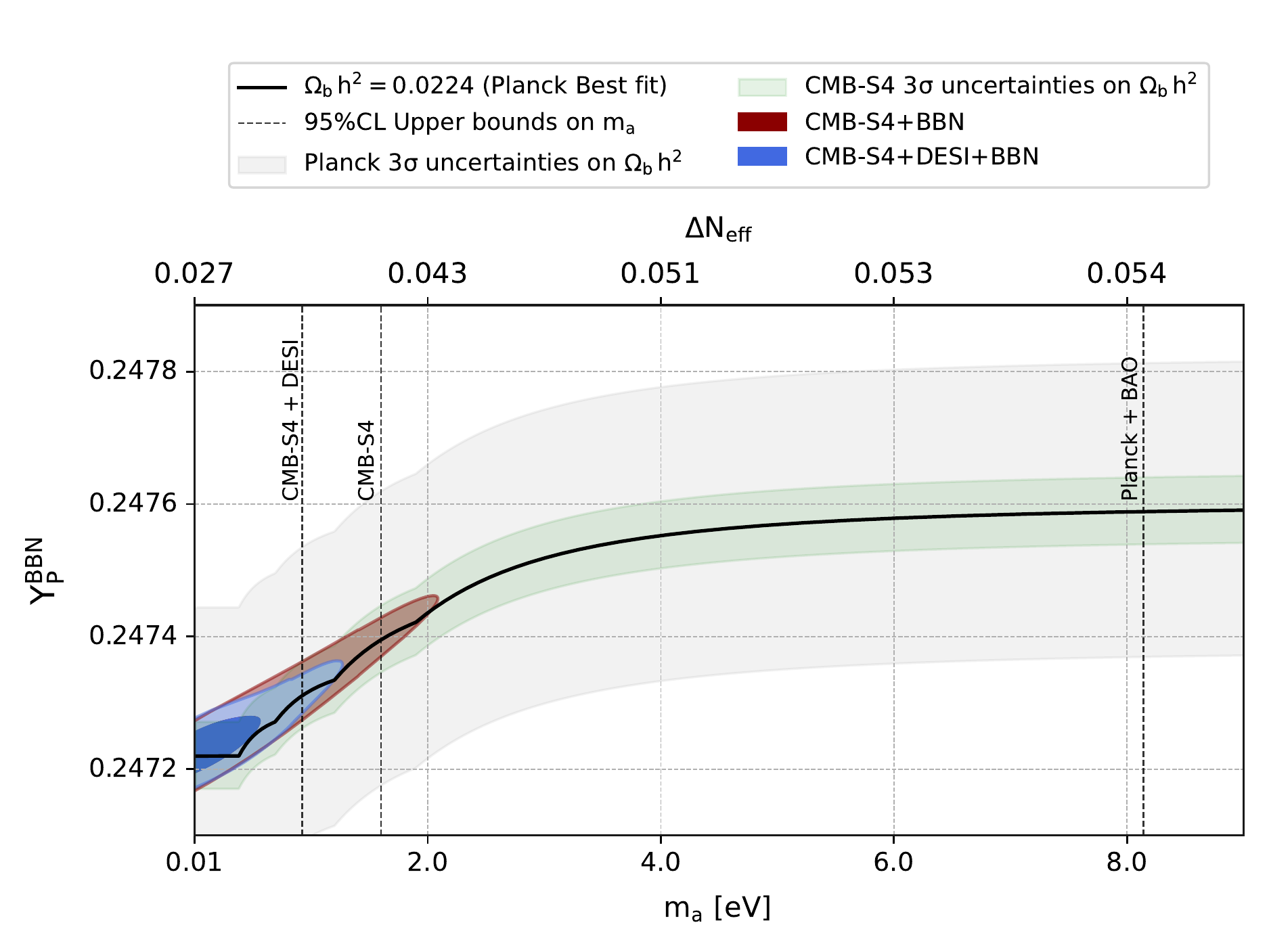}
\caption{Theoretical Helium fraction predictions in the $\Lambda\rm{CDM} + m_{\rm a} +\sum m_{\nu}$ cosmological model. The black solid line represents the Helium fraction as a function of the axion mass (with the corresponding $\Delta N_{\rm eff}$ on the top axis) obtained fixing $\Omega_b\,h^2=0.0224$. The green (gray) region represents the $3\sigma$ uncertainties on $Y_p^{\rm BBN}$ by CMB-S4 (Planck). The vertical lines are the 95\% CL upper limits on $m_{\rm a}$ from current cosmological data \citep{Giare:2020vzo} and from CMB and BAO future experiments obtained in this work, together with the respective 68\% and 95\% CL contours.}
\label{fig:Yp_DNeff_ma}
\end{figure*}

\section{Results}\label{Sec.Results}

In this section we discuss the results obtained with the forecasting method presented in the previous section. As a baseline model, we employ an extension of the standard cosmological model that includes both neutrinos and axions as thermal massive relics. 
We refer to it with $\Lambda\rm{CDM} + \sum m_{\nu} + m_a$. Within this model, we study the improvement on the bounds of QCD axions achievable by future CMB and BAO experiments. As aforementioned, thermal axions also contribute as additional relativistic species prior to recombination, increasing the value of $N_{\rm eff}$ thus leading to modification of the standard BBN predictions. Therefore we also take into account the effect of additional thermal species on the observational predictions of BBN light element abundances. Finally, we shall also compare the constraints on the hot relics and on the Helium nucleon fraction $Y_p^{\rm BBN}$ achievable with our simulated datasets without employing the BBN code, testing the dependence of our results on the assumptions adopted for the BBN sector and proving their robustness.

\subsection{Mixed Hot Dark Matter: Axions and Neutrinos}

\autoref{tab.Results.HDM} summarizes the results obtained from our forecasting methods for future CMB and BAO experiments while \autoref{fig:LCDM+ma+mnu} shows the $68\%$ and $95\%$~CL contour plots for different cosmological parameters. 

Using our forecasting data for future CMB-S4 observations, we derive the 95\% CL upper bounds on thermal relics of $\sum m_{\nu}<0.183\,\rm{eV}$ and $m_a<1.60\,\rm{eV}$. These values should be compared with those derived in~\citep{Giare:2020vzo} for the same cosmological model, exploiting the last CMB data release provided by the Planck Collaboration. In particular, one can appreciate that future CMB-S4 measurements are expected to improve the current bounds on the axion-gluon interaction scenario by a factor of $\sim 5$, while we estimate the improvement in the constraining power on the neutrino sector to be $\sim 2$. This enhancement in the constraining power on thermal relic masses is mostly due to the much higher sensitivity to the effective number or relativistic degrees of freedom $N_{\rm eff}$ expected from future CMB measurements~\footnote{We recall that, while the current Planck data leads to a 95\% CL upper limit of $\Delta N_{\rm eff}\lesssim 0.4$, future CMB-S4-like experiments are expected to bring this upper limit down by a factor of $\sim 10$, resulting in a much more tighter limit on dark radiation, $\Delta N_{\rm eff}\lesssim 0.06$ and eV-scale light relics \citep{DePorzio:2020wcz}.}.

Notice that, due to the degeneracy between the axion and the neutrino masses discussed in the introductory section, the contours in the ($\sum m_\nu$, $m_{\rm a}$) show a clear anti-correlation. Furthermore, these two parameters show very similar degeneracies with other cosmological parameters such as $H_0$, $\sigma_8$ and $\Omega_m$. It is well-known that hot thermal particles suppress structure formation at small scales and therefore galaxy clustering information becomes crucial to set bounds on the amount of dark matter in the form of these relics. As discussed in~\citep{Giare:2020vzo} the largest impact on CMB bounds on hot relics arises from the inclusion of the large-scale structure information from BAO measurements. For this reason, here, together with the likelihood for future CMB-S4 observations, we consider also a likelihood for future BAO measurements from the DESI-like experiment. Combining our simulated CMB-S4 and DESI forecasts, we obtain a further improvement in the cosmological constraining power for thermal relics, reaching the 95\% CL limits $m_{\rm a}<0.924\,\rm{eV}$ and $\sum m_{\nu}<0.122\,\rm{eV}$. In this case these bounds can be compared with those obtained for current Planck+BAO real data~\citep{Giare:2020vzo}, observing an improvement of a factor $\sim 8$ and $\sim 1.5$ in the sensitivity to the axion and neutrino masses, respectively. 

Our results clearly state that future cosmological observations can substantially improve the current constraints on $m_{\rm a}$, exploring a much larger range of the parameter space currently allowed for QCD thermal axion and reaching the sub-eV mass range. Conversely, when axions are included in the picture as additional thermal species, the possibility to detect the expected minimum neutrino mass of $0.06$~eV is no longer possible, and only upper bounds,  close to the inverted mass ordering prediction, can be derived.

\subsection{Primordial abundances of light elements}

\begin{figure}
    \centering
    \includegraphics[width=0.48\textwidth]{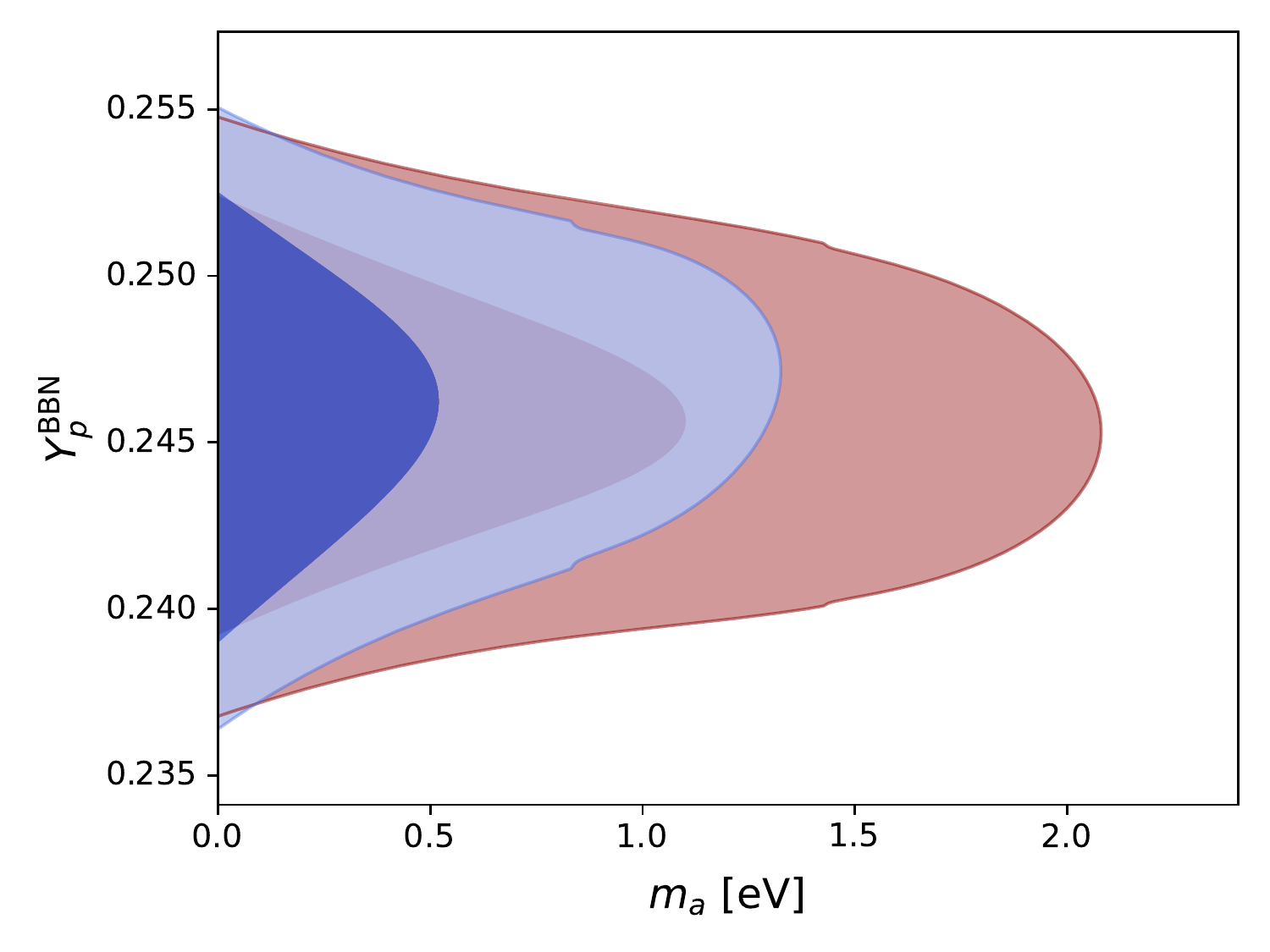}
\caption{Marginalized 2D posteriors in the plane $(m_{\rm a}\,,\,Y_p^{\rm BBN})$ when the BBN predictions are relaxed and the Helium fraction is considered as a free parameter.}
\label{fig:ma_Yp}
\end{figure}

Thermal axions contribute to the effective number of relativistic degrees of freedom, modifying the expansion rate at the radiation epoch and affecting, indirectly, the canonical BBN predictions. Even tough the latest results of the Planck collaboration place tight bounds on both the baryon density ($\Omega_b h^2 = 0.0224 \pm 0.0001$) and $N_{\rm{eff}}$ (limiting the amount of  additional relativistic degrees of freedom to $\Delta N_{\rm eff}\lesssim 0.4$), the impact of axions on the Helium fraction is extremely small and the Planck uncertainties on $\Omega_b\,h^2$ are still too large to provide robust theoretical predictions on the Helium abundance in presence of the axion. However, the next generation of CMB and BAO observations will substantially improve the bounds on the baryon energy density by a factor of 2, strongly reducing the theoretical uncertainties on $Y_p^{\rm BBN}$, and, possibly, allowing to test signatures of the axion in the primordial abundances. \autoref{fig:Yp_DNeff_ma} shows a comparison between CMB-S4 and Planck in determining the Helium fraction. In particular, we show the theoretical Helium fraction predictions in the $\Lambda\rm{CDM} + m_{\rm a} +\sum m_{\nu}$ cosmological model as a function of the axion mass (or, equivalently, as a function of the axion contribution to $\Delta N_{\rm eff}$) together with the 2D marginalized posterior distribution obtained for the CMB-S4 and CMB-S4+DESI simulated data. Notice that the BBN predictions introduce a strong correlation between the axion mass and the Helium fraction ($Y_p^{\rm BBN}$) that, combined with the substantial improvement in the constraining power expected by CMB-S4 and DESI, suggests that the BBN could be a useful tool to make predictions on hot axions and that astrophysical measurements of the primordial fraction of Helium could be used as independent test together with the cosmological observations. For this reason we included all the BBN light elements in our analysis. We provide the 68 \% CL results for the other light elements up to Beryllium-7 in~\autoref{tab.Results.HDM}. It should be noticed however that these results are derived without considering the experimental error in the measurement of the neutron life-time $\tau_n$. This error could dominate the total error budget, enlarging the theoretical uncertainties on the BBN predictions for $Y_p^{\rm BBN}$ by a factor of $\sim 3$ ($\Delta Y_p^{\rm BBN}(\Delta \tau_n)\simeq 0.00012$), producing the same effect as an extra dark radiation component~\citep{Capparelli:2017tyx}. Consequently a large degeneracy between the axion mass and the neutron life-time is expected and this effect may change the correlations between the primordial Helium fraction and the axion mass. For this reason, to prove the robustness of our results on hot massive relics, it is mandatory to follow also a very conservative approach and study the impact of additional hot relics on the abundances of primordial elements without assuming the BBN theoretical predictions but leaving all the parameters varying in uninformative flat priors. We therefore analyze a cosmological model where, together with axions and massive neutrinos, we also include the abundance of primordial Helium as an additional free parameter. We refer to this model as $\Lambda$CDM+$m_{\rm a}$+$\sum m_{\nu}$+$Y_p^{\rm BBN}$ and report the results obtained with our CMB and BAO forecasting data in \autoref{tab.Results.LCDM+ma+mnu+Yp}. In this case, the 68\% and 95\% CL marginalized contours in the plane ($m_{\rm a}$ , $Y_p^{\rm BBN}$) are shown in \autoref{fig:ma_Yp}. Removing the BBN predictions the strong positive correlation between the axion mass and the Helium fraction $Y_p^{\rm BBN}$ is relaxed as well. Furthermore the bounds on the Helium fraction are much less constraining, with 68\% CL bounds of $Y_p^{\rm BBN}=0.2458^{+0.0057}_{-0.0058}$ and $Y_p^{\rm BBN}=0.2460^{+0.0057}_{-0.0058}$ for CMB-S4 and CMB-S4+DESI, respectively.

On the other hand, the constraints on hot dark matter are basically unchanged. Exploiting our forecasting data for future CMB-S4 observations we can still derive the 95\% CL upper bounds $m_{\rm a}<1.63$ eV and $\sum m_{\nu}<0.183$~eV for axions and neutrinos, respectively. The upper limit on the total neutrino mass is exactly the same as that derived including the BBN code as well as the upper bound on the total axion mass. Similarly, combining future CMB-S4 and BAO data the upper bound on neutrinos masses is unchanged ($\sum m_{\nu}<0.120$ eV at 95\% CL) while the upper bound on axions is only slightly worsened to $m_{\rm a}<0.991$ eV at 95\%CL. These results prove that the impact of the BBN uncertainties on axion and neutrino masses is negligible and therefore the extraction of both $m_{\rm a}$  and $m_{\nu}$ does not rely on the assumptions adopted for the neutron-life time.

\section{Discussions and Conclusions} \label{Sec.Conclusions}
One of the declared targets of the next generation CMB and BAO cosmic observers is to improve the current constraining power on the neutrino sector, reaching a much better sensitivity both on their total mass and on extra dark radiation, severely constraining additional contribution to the number of relativistic degrees of freedom in the early Universe. In this study we have translated the expected experimental improvements into constraining power for well motivated physical extensions of the Standard Model of elementary particles, involving QCD axions as the solution to the strong CP problem. In particular, we have focused on axions produced thermally before the QCD phase transition via scattering with free gluons and analyzed mixed hot dark matter scenarios that include also neutrinos as additional massive relics. Assuming a fiducial $\Lambda$CDM cosmological model and following robust numerical procedures, we have simulated forecasting data for future cosmic microwave background (CMB) and baryon acoustic oscillation (BAO) observations, focusing in particular on CMB-S4-like and DESI-like experiments. Exploiting these simulated data we have studied the improvements in the cosmological constraints on axion and neutrino mixed hot dark matter scenarios. We have shown that future CMB-S4 measurements are expected to improve the current cosmological bounds on axion-gluon interactions by a factor of $\sim 5$. The improvement in the constraining powering on the neutrino sector is $\sim 2$, and the expected 95\% CL upper bounds on hot relic masses are $\sum m_{\nu}<0.183\,\rm{eV}$ and $m_{\rm a}<1.60\,\rm{eV}$. Since hot thermal particles such as axions and neutrinos suppress structure formation at small scales, galaxy clustering information becomes crucial to set bounds on the amount of hot dark matter in the form of thermal relics. For this reason we combine the CMB-S4 simulated data with our likelihood for future BAO DESI-like experiments, showing that in this case the bounds on thermal relics can be further improved to $m_{\rm a}<0.924\,\rm{eV}$ and $\sum m_{\nu}<0.122\,\rm{eV}$, both at 95\% CL. Interestingly, future cosmic observers can substantially improve the current constraints on $m_{\rm a}$, reaching the sub-eV mass range. Conversely, when axions are included in the picture as additional thermal species, a  two sigma detection of the neutrino mass for a fiducial value of $\sum m_{\nu}=0.06\,\rm{eV}$ (which is another declared target of the next-gen CMB and BAO experiments) is excluded. Only upper bounds (close to the inverted mass ordering prediction) can be derived in the former case. Finally, since thermal axions also contribute as additional relativistic species prior to recombination increasing the value of $N_{\rm eff}$ and thus leading to a modification of the standard Big Bang Nucleosynthesis (BBN) predictions, we have also taken into account the effect of an additional thermal species on the observational prediction of BBN on light element abundances up to Beryllium-7. We have compared the constraints on hot relics achievable within our simulated datasets with and without employing the \textsc{PArthENoPE} BBN code for computing the theoretical predictions of the different abundances of primordial elements. We have shown that our results do not rely strongly on the assumptions adopted in the BBN sector, definitively proving their robustness. We conclude supporting and underlying the relevance of multi-messengers searches of axions, neutrinos and primordial  light element measurements. Indeed, cosmology-independent limits on the axion and neutrino masses, combined with precise astrophysical measurements of light elements, may provide an important cosmological test for checking the BBN predictions. On the other hand, future cosmic observations should also be able to probe scenarios with hot axions with masses $m_{\rm a}\gtrsim 1$~eV and a missing evidence would constrain the axion mass at the sub-eV level, favoring the normal ordering as the one governing the mass pattern of neutral fermions. In the same multi-messenger spirit, future cosmology-independent probes of neutrino masses (i.e.  future terrestrial double beta decay and/or long baseline neutrino experiments) will play, even if indirectly, a crucial role on axion searches. Complementarity between astrophysical and cosmological axion searches have been carefully explored in the very recent study of Ref.~\citep{Green:2021hjh}, which appeared while this manuscript was being completed. Our results agree fairly well with those obtained by the authors of the reference above.

\section*{Acknowledgments}
We thank Francesco D'Eramo, Seokhoon Yun, Sadra Hajkarim and Eoin O Colgain for useful comments and suggestions.

W.G. and A.M. are supported by "Theoretical astroparticle Physics" (TASP), iniziativa specifica INFN. F.R. acknowledges support from the NWO and the Dutch Ministry of Education, Culture and Science (OCW), and from the D-ITP consortium, a program of the NWO that is funded by the OCW. 
OM is supported by the Spanish grants FPA2017-85985-P, PROMETEO/2019/083 and by the European ITN project HIDDeN (H2020-MSCA-ITN-2019//860881-HIDDeN).
EDV acknowledges the support of the Addison-Wheeler Fellowship awarded by the Institute of Advanced Study at Durham University.
In this work we made use of the following \texttt{python} packages that are not mentioned in the text: \texttt{SciPy}~\citep{2020SciPy-NMeth} for numerical sampling of the statistical distributions involved in our data analysis, \texttt{GetDist}~\citep{Lewis:2019xzd} a tool for the analysis of MCMC samples which employs \texttt{Matplotlib}~\citep{Matplotlib} for the realization of the plots in the paper and \texttt{NumPy}~\citep{NumPy} for numerical linear algebra. 
\section*{Data Availability}
The simulated data underlying this article will be shared on reasonable request to the corresponding author.



\bibliographystyle{mnras}
\bibliography{MNRAS} 



\appendix

\section{Degrees of freedom in the Early Universe}
\label{AppendixA}

The results derived in this paper, assume a minimal extension of the Standard Model that includes QCD axions thermalized at high temperatures, prior to the QCD phase transition. However, little is known about the number of degrees of freedom above the QCD epoch and particularly around the Electroweak transition.  In many extensions of the Standard Model of particles, such as supersymmetry, additional degrees of freedom in the Early Universe are in principle allowed, possibly changing the contributions of QCD axions to the different cosmological observables and, consequently, our forecasted bounds.

In this appendix, we briefly discuss what happens to our forecasted limits in many-degree-of-freedom models of  the Early Universe. In particular, as a toy model, we consider  Minimal Supersymmetrical extensions of the Standard Model (MSSM) \cite{Ghodbane:2002kg} where, above the Electroweak scale, the relativistic degrees of freedom are more than doubled since all supersymmetric partners can eventually be excited. More precisely, when the temperature is larger than all particle masses, within the MSSM we have $g_{\star}^{\rm MSSM} = 228.75$ \citep{Schwarz:2003du}, while within the SM we have $g_{\star}^{\rm SM} = 106.75$).

\begin{figure}[htb]
    \centering
    \includegraphics[width=\columnwidth]{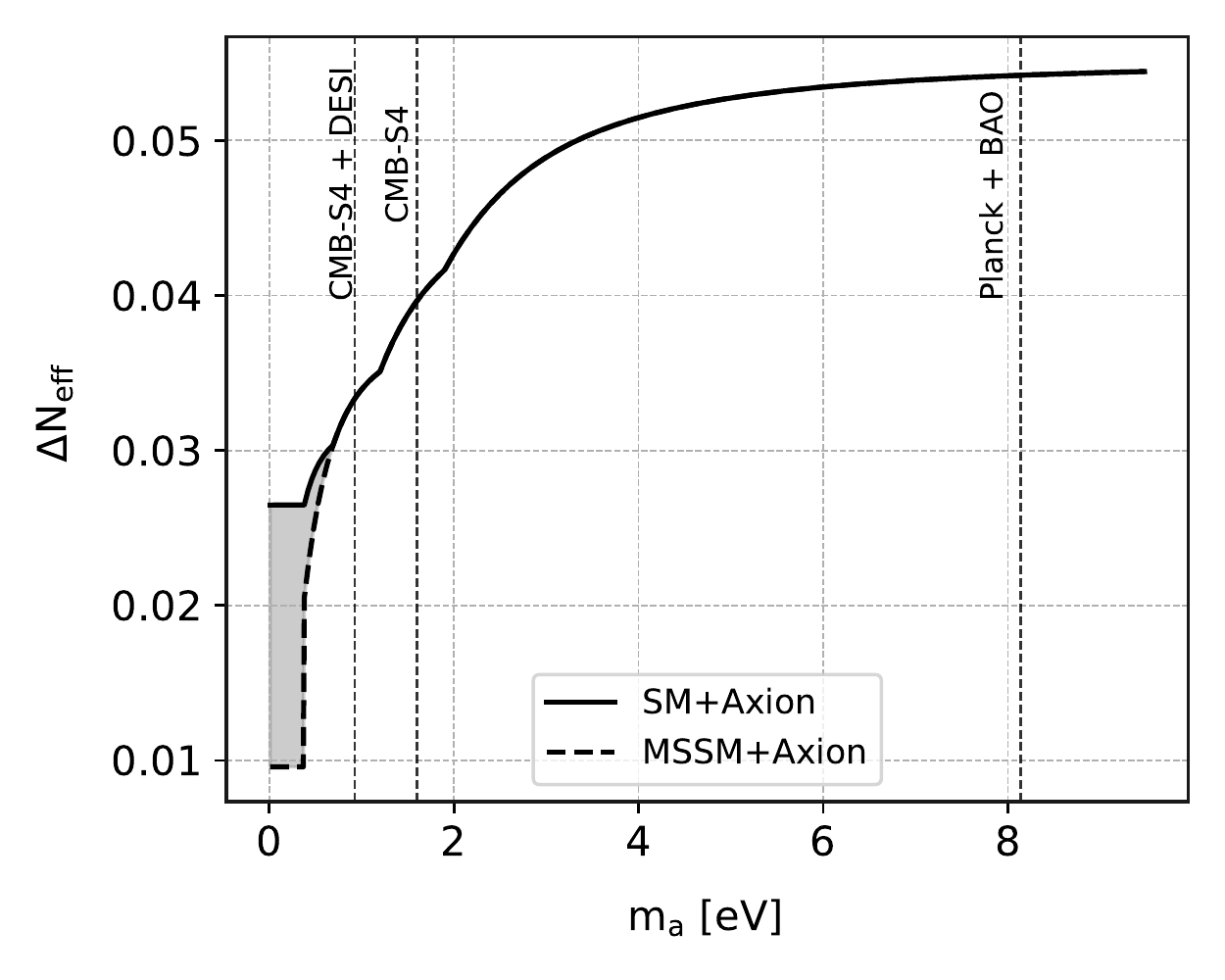}
\caption{Axion contribution to $\Delta N_{\rm eff}$ in a minimal supersymmetrical extension of the Standard Model.}
\label{fig:susy}
\end{figure}

Retracing our analysis in such scenarios, it is clear that, if the axion decouples from the thermal bath at sufficiently high temperatures, its residual number density is suppressed by the additional entropic degrees of freedom and consequently also its contribution to the effective number of relativistic degrees of freedom will be smaller. However, although the supersymmetric model considered here has light superpartners compared with the other models proposed in literature, the lightest supersymmetric particle in the picture is the lighter neutralino with a mass of $\sim 95$ GeV. Therefore, the additional supersymmetric degrees of freedom can be excited only at temperatures of the order of the EW scale ($T\gtrsim 100$ GeV) that, by Eq.~\eqref{Td_fa}, correspond to tiny axion masses $m_a\lesssim 0.5$~eV. As shown in \autoref{fig:susy}, these mass values are beyond (even though close) the sensitivity expected by future cosmological and astrophysical experiments. Nonetheless, these effects can be relevant to the abundance of primordial Helium predicted by the BBN. Indeed, in the limit of high decoupling temperatures ($T\gg 100$ GeV), the contribution of the axion to the effective number of relativistic degrees of freedom is reduced from $\Delta N_{\rm eff}\simeq 0.027$ to $\Delta N_{\rm eff}\lesssim 0.01$, see also \autoref{fig:susy}. This difference changes the theoretical predictions shown in \autoref{fig:Yp_DNeff_ma} for $Y^{\rm BBN}_{P}$ vs $m_a$. Therefore, we recommend caution when applying our forecasted limits on primordial abundances to scenarios that involve additional degrees of freedom in the Early Universe.

\bsp	
\label{lastpage}
\end{document}